\documentclass[showpacs,showkeys,twocolumn,prl,preprintnumbers,amsmath,amssymb]{revtex4}
\newcommand{\beq}{\begin{equation}}
\newcommand{\eeq}{\end{equation}}

\newcommand{\id}{i\kern.06em\hbox{\raise.25ex\hbox{$/$}\kern-.60em$\partial$}}

\newcommand{\bs}{/\kern-.52em b}
\newcommand{\qs}{/\kern-.52em s}

\newcommand{\yp}{^{\prime}}

\newcommand{\dd}
{\kern.06em\hbox{\raise.25ex\hbox{$/$}\kern-.60em$\partial$}}

\newcommand{\vep}{\varepsilon}

\newcommand{\bi}{{\bf i}}
\newcommand{\bj}{{\bf j}}

\newcommand{\bk}{{\bf k}}

\newcommand{\JK}{J_{\rm K}}

\newcommand{\bS}{{\bf S}}

\newcommand{\bra}[1]{\langle #1|}
\newcommand{\ket}[1]{|#1\rangle}

\newcommand{\lj}{\langle}
\newcommand{\rj}{\rangle}
\DeclareMathAlphabet{\mathpzc}{OT1}{pzc}{m}{it}
\usepackage{mathrsfs}
\input amssym.def
\input amssym
\input epsf
\usepackage{graphicx}
\usepackage{dcolumn}
\usepackage{bm}
\usepackage{epsfig}
\begin{document}
\title{Ferromagnetism of Ga$_{1-x}$Mn$_x$As and Weiss theory of
           Curie temperature in the coherent potential approximation
        }
\author{Sze-Shiang Feng$^\dag$, Mogus Mochena}
\affiliation
{Physics Department, Florida A \& M  University, Tallahassee, FL 32307}

\date{\today}
\begin{abstract}

The zinc-blende GaAs-based  III-V diluted magnetic
semiconductors (DMS)are studied in the coherent potential
approximation(CPA). In this work, we use the exact Hilbert transformation of the
face-centered cubic(fcc) density of states (DOS), which is different from the usual semi-circle density
of states employed in our previous work. Our calculated
relation of ground-state energy and impurity magnetization shows
that ferromagnetism is always favorable at low temperatures. For
very weak Kondo coupling, the density of states (DOS) of the host
semiconductor is modified slightly. Impurity band can be generated
at the host band bottom only when Kondo coupling is strong enough.
Using Weiss molecular theory, we predict a nonlinear relation of
Curie temperature with respect to Kondo coupling, as is different from the conclusion of our previous calculations
based on semicircle DOS.  The agreement of our calculated $T_{\rm C}$
with measured values is convincing.
\end{abstract}
\pacs{75.30.Ds, 75.50.Pp,75.50.Dd}
\keywords{Ferromagnetism, Curie temperature, coherent potential approximation, Ga$_{1-x}$Mn$_x$As}
\maketitle The ferromagnetism of DMS of III-V-type is not well
understood. To explain ferromagnetism in DMS, various models and
approaches have been
proposed\cite{Oiwa:1997}-\cite{Rajagopal:1998}. Though the models
differ from each other in details, they all agree that the
coupling between the carriers and local spins is of fundamental
importance. An issue of debate, however, is how the exchange
between localized spins is induced by the carriers. One model for
this induced exchange is the Ruderman-Kitttel-Kasuya-Yosida (RKKY)
interaction\cite{Oiwa:1997}\cite{Dietl:1997}. Another version
which results in conclusion equivalent to RKKY
 is the Zener model\cite{Dietl:2000&2001} which uses the fact that the valence holes are on $p$-orbitals.
 A third model is the double-exchange (DE) mechanism\cite{Akai:1998}. But this model
is inconsistent with the charge-transfer
 properties\cite{Dietl:2000&2001}.
Though RKKY can give a Curie temperature in agreement with
experiment, some argue that the RKKY model breaks down
here\cite{Konig:2000}\cite{Matsuda:2004} because the local coupling between the
carrier and the impurity spin is much larger than the Fermi energy
and can not be treated perturbatively. In dealing with the effect
of the localized spins, a key issue is whether or not randomness
should be taken into consideration. The above models are all mean
field approximations (MFA) which assume homogeneity and neglect
randomness. But DMSs are disordered systems with positional
disorder of Mn impurities. As concluded in\cite{Dietl:1997}\cite{Timm:2003}, disorder has
a substantial influence upon carrier magnetic susceptibility.
Hence, any first principle consideration should take into account
the randomness of the
impurities.\\
\indent A classic method of dealing with randomness is the
coherent potential approximation(CPA)\cite{Soven:1967} which has
been applied to
DMS\cite{Takahashi:1999&2001}-\cite{Bouzerar:2003}. Basing on the
formalism of\cite{Rangette:1973} and assuming very large local
spin $S$ while keeping the product $IS$ constant (where $I$ is the
Kondo -like interaction), \cite{Takahashi:1999&2001} obtained the
density of states and the relation between Curie temperature and
the doping concentration. Using the averaged carrier Green's
function, \cite{Bouzerar:2003} arrived at the conclusion that the
local coupling between the carrier and the impurity spins must be
intermediate in order to acquire ferromagnetism.
In our previous work\cite{Feng:2005},
 we
used the formalism of CPA in \cite{Takahashi:1996} to study the
ground-state properties of III-V DMS. The DOS for the undoped crystal
we used is the semi-circle DOS. In contrast to
\cite{Takahashi:2003}, we kept $S=5/2$ and treated the impurity
spins fully quantum mechanically. Though it is mostly accepted
that in III-V type DMS the effective spin of valence holes is
$3/2$\cite{Linnarsson:1997}-\cite{Fiete:2003}, we described
the holes as spin-$1/2$ fermions. It is usually believed that such
a description can still catch the essential physics. Because of
spin-orbit interaction, the $p$-orbitals are spilt (with split-off
$\simeq 0.34$eV)  into a spin-$3/2$ multiplet and a spin-$1/2$
multiplet\cite{Luttinger:1955}. Using spherical approximation, the
kinetic energy of the Luttinger-Kohn Hamiltonian
\cite{Luttinger:1955} for the spin-3/2 multiplet takes the form
$\sum_\mu (\hbar^2\bk^2/2m_\mu)c^\dag_{\bk\mu}c_{\bk\mu}$ near the
valence top after diagonalization, where $m_\mu=m_h\simeq 0.5m $
for $\mu=\pm 3/2$ and $m_\mu=m_\ell\simeq 0.07m $ for $\mu=\pm
1/2$ ($m$ is the effective mass of a free hole ). The interaction
between the spin of holes and local 5/2-spins now takes a
$\bk$-dependent form $\sum_{\bi,\bk,\bk\yp}\bS_\bi\cdot c^\dag_\bk
{\bf J}(\bk,\bk\yp)c_{\bk\yp}\exp(-i(\bk-\bk\yp)\cdot{\bf R}_\bi)
$\cite{Zarand:2002}. Since the DOS for parabolic band is
$g(\vep)=(1/2\pi^2 \hbar^3)(2m)^{3/2}\sqrt{\vep}$, we have the
ratio of DOS $g_h(\vep)/g_\ell(\vep)\simeq 19$ for heavy holes and
light holes, i.e., most of valence holes are heavy holes.
Therefore, it is a valid approximation to consider only heavy
holes. Moreover, since the hole density is very small, the
Fermi wave vector is supposed to be very small and the it is thus
a reasonable approximation to consider ${\bf J}(\bk,\bk\yp)\simeq
{\bf J}(0,0)$ in the interaction term. And this leads to the usual
assumption that the carriers are shallow holes and the coupling of
the shallow holes to the Mn$^{2+}$ can be
described by local Kondo interaction between 5/2-spins and 1/2-spins.
 The system is highly compensated\cite{Matsukura:1998}-\cite{Beschoten:1999}
and the carriers are holes originating from randomly distributed Mn.
There are
different kinds of randomness, e.g. substitutional randomness,
interstitial randomness, antisite randomness, and directional
 randomness of impurity spin. It is commonly agreed now that interstitial Mn atoms and
 antisite As only reduce the hole densities and
 do not affect conduction of holes significantly. Therefore we consider only two
kinds of randomness, i.e., the random substitution of the Mn atoms
and the random direction of the impurity spins.
\\
\indent  Since the DOS for undoped crystal is much involved in the CPA calculations and the real structure
of Ga$_{1-x}$Mn$_x$, where the doping
concentration $x$ varies from 0.015 to 0.08 in region of interest
for ferromagnetism\cite{Edmonds:2002}, is of zinc-blende structure, we here use the method developed
 in our previous work
to study DMS Ga$_{1-x}$Mn$_x$As, taking into consideration the real fcc lattice structure.
In as-grown samples of \cite{Edmonds:2002}, ferromagnetism
was realized at Curie temperature of 76K at $x=0.05$. The hole
density $p$ varies from $5.6\%x\sim 93\%x$.  We use the
tight-binding model Hamiltonian \beq
H=\sum_{\bi,\bj,\sigma}t_{\bi\bj}c^\dag_{\bi\sigma}c_{\bj\sigma}+\sum_\bi
u_\bi \eeq where $u_\bi$ depends on whether $\bi$ is a Ga or Mn
site. For Ga-site $u_\bi=u^{\text{G}}_\bi=E_{\text{G}}\sum_\sigma
c^\dag_{\bi\sigma}c_{\bi\sigma}$ , and for Mn-site
$u_\bi=u^{\text{M}}_\bi=E_{\text{M}}\sum_\sigma
c^\dag_{\bi\sigma}c_{\bi\sigma}+\JK\bS_\bi\cdot{\bf s}_\bi$.
$\bS_\bi$ is the local spin of Mn at site $\bi$, ${\bf
s}=(1/2)c^\dag_\sigma\mbox{\boldmath$\tau$}_{\sigma\sigma\yp}c_{\sigma\yp}$
is the spin of a hole where  $c^\dag_\sigma(c_\sigma)$ is the
creation(annihilation) operators for holes , spin indices
$\sigma,\sigma\yp=\uparrow,\downarrow$, and
$\mbox{\boldmath$\tau$} =(\tau_1, \tau_2, \tau_3)$ are the three
usual Pauli matrices. $E_{\rm M}$ and $E_{\rm G}$ are the on-site
energies for Ga and Mn and are assumed constant. The hopping
energy $t_{\bi\bj}=t$ if $\bi,\bj$ are nearest neighbors and zero
otherwise and $\JK>0$ is the local Kondo coupling.
  According to the general scheme of CPA, the virtual
unperturbed Hamiltonian is
\beq
\mathscr{H}(\vep)=\sum_{\sigma,\bk}(t_\bk+\Sigma_\sigma(\vep))c^\dag_{\bk\sigma}c_{\bk
\sigma}
\eeq
where $\vep$ is the Fourier frequency variable,
$\Sigma_\sigma(\vep)$ is the CPA self-energy to be determined
self-consistently and $t_\bk$ is the Fourier transformation of
$t_{\bi\bj}$. Then the relative perturbation $V$ is given by
\beq
V=H-\mathscr{H}(\vep)=\sum_\bi v_\bi
\eeq
where $v_\bi=v^{\rm G}_\bi=\sum_\sigma(E_{\text{G}}-\Sigma_\sigma)c^\dag_{\bi\sigma}c_{\bi
\sigma} $ for Ga and $v_\bi=v^{\rm
M}_\bi=\sum_\sigma(E_{\text{M}}-\Sigma_\sigma)c^\dag_{\bi\sigma}
c_{\bi \sigma}+\JK\bS_\bi\cdot{\bf s}_\bi $ for Mn. The reference
Green's function is $
\bra{\bi\sigma}\mathscr{R}(\vep)\ket{\bj\sigma\yp}=\bra{0}
c_{\bi\sigma}(\vep-\mathscr{H})^{-1}c^\dag_{\bj\sigma\yp}\ket{0} $
where $\ket{0}$ is the vacuum state of the $c$-operators, and the
associated $t$-matrices are $ t^{\rm G}_\bi=v_\bi^{\rm G}/(1-\mathscr{R}v_\bi^{\rm G})
 , t^{\rm M}_\bi=v^{\rm M}_\bi/(1-\mathscr{R}v^{\rm M}_\bi) $.  So the CPA equation
and DOS are given by
\beq
(1-x) t^{\rm G}_\bi +x\lj
t_\bi^M\rj_{\rm spin}=0 \label{CPA}
\eeq
and
\beq
g_\sigma(\vep)=-\frac{1}{\pi}{\rm Im} F_\sigma(\vep)
\eeq
where
$\lj\cdots\rj_{\rm spin}$ denotes average over the configurations
of impurity spins and $F_\sigma(\vep)$ is the Hilbert transformation of DOS,
$F_\sigma(\vep)=\bra{\sigma\bi}\mathscr{R}\ket{\bi\sigma}=(1/N)\sum_\bk
[1/(\vep-t_\bk-\Sigma_\sigma)] $, where $N$ is the number of
lattice sites. For any $f(S^z)$, the spin average is given by
$\lj f(S^z)\rj_{\rm spin}=\sum_{S^z=-S}^{S}e^{\lambda S^z} f(S^z)/\sum_{S^z=-S}^{S}e^{\lambda S^z}$ where
$\lambda$ is determined by the condition
$\lj S^z\rj_{\rm spin}=m$, $m$ is the given magnetization of the impurity spins.
In our single-particle CPA, the Callen-Shtrikman relation\cite{Callen:1965} that tells
 there is a one-to-one correspondence between $m$ and $\lj (S^z)^n\rj_{\rm spin}$ for $n>1$ applies .
These explicit forms of CPA equations (\ref{CPA}) are given by \cite{Takahashi:1996} in another context.
To solve CPA equations, we need to know $F^{(0)}_\sigma(\vep)$.
Jelitto\cite{Jelitto:1969} and Iwata \cite{Iwata:1969} first gave a rigorous calculation
  of the $\bk$-integrated Green's function for fcc tight-binding Hamitonian. The expression entails complete elliptic
  integral and the analytic properties
   are very complicated.  Morita\cite{Morita:1971} et al made further
  discussions later. A much improved expression of the $\bk$-integrated Green's function is obtained quite
  recently\cite{Joyce:2004}. In unit of $w/4$ where $w$ is the band width and defining
\beq
 \kappa(\vep)=\frac{1}{2}[1-\frac{4\sqrt{-\vep}\sqrt{1-\vep}-(1+\vep)\sqrt{1-\vep}\sqrt{-3-\vep}}{(1-\vep)^2}]
\eeq
then  $F^{(0)}(\vep)$ can be written as
\beq
F^{(0)}(\vep)=\frac{4}{\pi^2(1-\vep)^{3/2}}(\sqrt{-\vep-3}-2\sqrt{-\vep})\mathcal{K}^2(\kappa)
\eeq
where $\mathcal{K}(\kappa)$ is the complete elliptic integral
\beq
\mathcal{K}(\kappa)=\int^{\pi/2}_0 d\theta \frac{1}{\sqrt{1-\kappa\sin^2\theta}}
\eeq
At zero temperature, the carrier density for spin $\sigma$ can be
expressed as $
n_\sigma=\int^{\vep_F}_{-\infty}g_\sigma(\vep)d\vep,$
 where $\vep_F$ is the Fermi energy, and the total carrier density
$n=n_\uparrow+n_\downarrow$. The total electronic ground-state energy per site is $
\vep_g=\int^{\vep_F}_{-\infty}\vep
[g_\uparrow(\vep)+g_\downarrow(\vep)]d\vep $.\\
\begin{figure}
\epsfxsize= 9.4truecm {\epsfbox{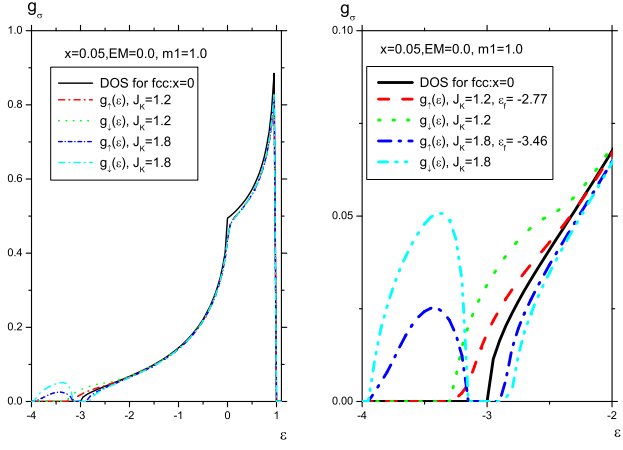}} \caption {Left
panel:DOS in the full range for some given model parameters;
Right panel: enlarged display from $\vep=-4$ to $\vep=-2$, the
Fermi energies are for as-grown samples from \cite{Edmonds:2002}}
\label{fig:g(z)}
\end{figure}
\indent In our numerical calculation,  we chose $w=4.0$ eV which
is roughly the bandwidth of the heavy holes\cite{Blakemore:1982}
and set $E_{\rm G}=0$ since we can shift the chemical potential
without loss of physics. (Note that the effective mass for
tight-binding spectrum is related to the bandwidth by
$m_h=8\hbar^2/(wa^2)$ where $a\simeq 5.6532$\AA. This relation
gives $m_h\simeq 0.48 m_e$ for $w=4$ eV). FIG.\ \ref{fig:g(z)}
shows DOS for some given model parameters. The corresponding hole
density is $p\simeq 0.442 x$ for as-grown samples as provided in
\cite{Edmonds:2002}. The left panel shows DOS in the full range
for some given model parameters; the right panel is an enlarged
display for the range $\vep=-4$ to $\vep=-2$. As a check for our
numerical results, the sum rule $\int ^{\infty}_{-\infty}
g_\sigma(z)dz=1$ and the relation $n_\uparrow+n_\downarrow=p$ are
also preserved where $n_\sigma$ is calculated from spin-resolved
DOS respectively. As can be seen that for $\JK=1.2$, there is no
separate impurity band and the
 DOS is not substantially
different from the bare DOS in shape. When Kondo coupling
$\JK=1.8$, separate impurity bands show up. Fig.1 also shows that
for $m>0$, there are always more spin-down carriers than spin-up
carriers, in compliance with the fact that the local $p$-$d$
coupling is antiferromagnetic\cite{Dietl:2000&2001}.
 \\
\begin{figure}
\epsfxsize= 9.4truecm {\epsfbox{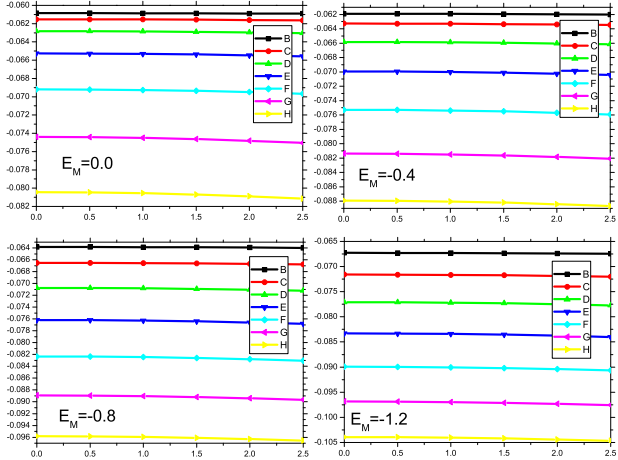}} \caption {
Relation of $\vep_g$ versus $m$. B: $\JK=0.6$; C: $\JK=0.8$; D:
$\JK=1.0$; E: $\JK=1.2$; F: $JK=1.4$; G: $\JK=1.6$; H: $\JK=1.8$.
$x=0.05, p=0.442x,w/4=1.0 $ eV} \label{fig:Eg(m,J,EM)_1}
\end{figure}
\begin{figure}
\epsfxsize= 9.4truecm {\epsfbox{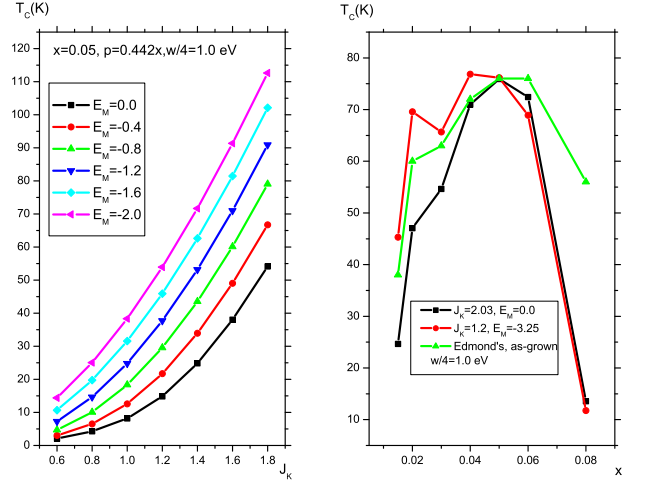}} \caption {Theoretical
estimation of Curie temperature using Weiss molecular theory}
\label{fig:Tc}
\end{figure}
\indent FIG.\ \ref{fig:Eg(m,J,EM)_1} shows the relation between ground-state energy
per site and the impurity magnetization $m$. In actual calculation of ground-state energy, an important issue
is the determination of the Fermi energy which is fixed by the integration of an interpolation function of the DOS. If the
interval $\delta\vep$ is not small enough, the integral of the interpolated DOS may vary significantly with the choice
of $\delta\vep$. To make it stable enough, we choose $\delta\vep=5\times 10^{-4}$ to interpolate DOS linearly.
Our results show
 that
for all the chosen values of model parameters, the ground-state energy per site always decreases, though very slowly, with increase
of impurity magnetization. Therefore CPA predicts that at very low temperatures,
 ferromagnetism is always energetically favorable for all the model parameters considered. \\
\indent In FIG.\ \ref{fig:Tc}, we show the dependence of Curie temperature on the model parameters and the doping concentration.
As proposed in our previous work \cite{Feng:2005},
Weiss molecular field theory can be employed to calculate the Curie temperature.
Given $m$, one can calculate DOS and then $\lj s_z\rj$. So one can establish
relation $\lj s_z\rj=\lj s_z\rj(m)$.
On the other hand , given $\lj s_z\rj$, each impurity spin feels an effective field $\JK\lj s_z\rj$ and thus we have
$ m=SB_S(\beta h) $
 with $h=-\JK\lj s_z\rj(m),\beta=1/k_{\rm B}T$ and $B_S(x)$ the conventional Brillouin function.
 For very small $m$, we have
 $
\lj s_z\rj\simeq-Am
$
with $A>0$ and we have
$
\beta h\simeq\beta \JK A m
$
. So
$
B_S(\beta h)\simeq (S+1)\beta h/3
$
and thus the Curie temperature can be estimated by
$
k_{\rm B}T_{\rm C}\simeq \JK S(S+1) A/3
$
 . For small $m$,  letting
$
F_\uparrow(z)=F(z)+\psi(z)m,
F_\downarrow(z)=F(z)-\psi(z)m
$
where $F(z)$ is the paramagnetic solution,  we have
\beq
A(\JK,E_{\rm M},x,\beta)\simeq \frac{1}{\pi}\int^{\vep_F}_{-\infty} d\vep\,\, {\rm Im}\,\psi(\vep)
\eeq
where we have ignored the $\beta$-dependence. As our numerical results (not shown here) indicate,
the chemical potential
is very close to the zero temperature Fermi
energy in a wide range of temperatures
($\beta> 50$), showing that the Fermi function can be approximated by the zero temperature step
function.
The left panel in FIG.3 shows the relation of $T_{\rm C}$ versus $\JK$ for different values
of $E_{\rm M}$. Unlike the result from semicircle DOS \cite{Feng:2005}, the curves exhibit a nonlinear relation.
The right panel shows the dependence of Curie temperature on the doping concentration for $\JK=2.03,E_{\rm M}=0$ and
$\JK=1.2,E_{\rm M}=-3.25$ for the as-grown samples in \cite{Edmonds:2002}. These two sets of parameter values
are chosen so that at $x=0.05, T_{\rm C}=76$ K. It is seen that our CPA+Weiss approach for $T_{\rm C}$ does reveal
the observed behavior. Our result show that both two sets of values $\JK, E_{\rm M}$ can explain the measured
results.
\\
\indent To conclude, we summarize our results here. Using CPA and treating the impurity spins fully
quantum mechanically, we have calculated the ground-state energies of GaAs-based III-V DMS for a wide range
of model parameters. The results show that ferromagnetism is always preferable at low temperatures
. With the help of Weiss molecular theory of ferromagnetism, we obtained a nonlinear relation of Curie temperature
with respect to Kondo coupling.  Our theoretical prediction of relation $T_{\rm C}\sim x$ does
fit quite well the measured results for as-grown samples \cite{Edmonds:2002}.
The deviations of our calculated results from the experimental values
is due to the sensitivity of $T_{\rm C}$ to the hole density or the ratio $p/x$, which
 is difficult to be measured precisely.
Though the value of $\JK=1.2$ eV is more generally accepted than other values, the actual
value of $\JK$ is still an issue of debate. Our calculation allows adjustment of model parameters $\JK, E_{\rm M}$
and the main features of numerical results for different model parameter values remain similar.
 Finally, a number of complicated models with complicated
 approximations have been proposed to calculate Curie temperatures (e.g.\cite{Bouzerar:2005}).
 Yet, we believe that CPA, a method having been tested for decades,
 is still an effective way to understand ferromagnetism in DMS.

The authors thank Prof. Rajagopal for his helpful discussions. We are grateful to Dr. Edmonds
                for providing additional data.
This work is supported in part by the Army High Performance Computing Research
Center(AHPCRC) under the auspices of the Department of the Army, Army Research Laboratory (ARL) under
Cooperative Agreement number DAAD19-01-2-0014.\\
$^\dag$Corresponding author, e-mail: shixiang.feng@famu.edu

\end{document}